# New Ghost States in SU(3) Gauge Field Theory

Scott Chapman
440 N Wolfe Rd.
Sunnyvale, CA 94085
scottc@cal.berkeley.edu

**Abstract**

The ghost sector of SU(3) gauge field theory is studied, and new BRST-invariant states are presented that do not have any analog in other SU(N) field theories. The new states come in either ghost doublets or triplets, and they appear exclusively in SU(3) due to the fact that the non-Abelian part of the BRST charge has 3 ghost operators, while SU(3) has 3 pairs of off-diagonal gauge constraints. The states have finite, positive norms even though the triplet states do not have well-defined ghost numbers. It is speculated that this special nature of the ghost sector of SU(3) could play some role in QCD confinement.

## Introduction

The mechanism behind quark confinement in QCD is still somewhat of an open question. So far, it has not been possible to derive confinement analytically from SU(3) guage field theory, so heuristic arguments, phenomenological models and numerical lattice methods have been employed instead. In this paper, some unique peculiarities of SU(3) quantization are explored in the hope that they will form a step toward gaining a deeper analytic understanding of the origin of quark confinement.

The most common way to quantize SU(3) and other non-Abelian gauge theories is to use the path-integral formalism with Fadeev-Popov ghosts. Many practitioners of this formalism point out that in order to put classical path integrals on the firmest theoretical foundation, they should first be derived from Hamiltonian canonical quantization [1,2]. For non-Abelian field theories, the BRST operator method is the most consistent and widely accepted form of Hamiltonian canonical quantization.

In the BRST quantization of gauge field theories, physical gauge invariance is enforced by requiring physical states to be BRST-invariant, which means that they are annihilated by the BRST charge. The BRST charge for a given theory is formed by multiplying the theory's gauge constraints by ghosts, then by including additional ghost terms that make the charge nilpotent. For non-Abelian field theories, the easiest BRST-invariant states to construct are those that are annihilated either by all of the gauge constraints (Dirac condition) or by all of the ghost operators. Both of these classes of states have zero ghost number (as long as the non-minimal sector and their anti-ghosts are included). They also both preserve the threefold symmetry of SU(3) by not treating any gauge "direction" differently from any other. Presented here are new classes of BRST-invariant states that do not have well-defined ghost number and that treat the various gauge "directions" differently.

In the first section, standard definitions are given, and the BRST-invariant states usually used in non-Abelian gauge theories are presented. This section also presents the method developed in [3-7] of using an indefinite metric in the non-minimal sector in order to create states with well-defined norms. In the second section, new notation is defined and the new solutions are presented. The last section contains summary comments and new questions.

## Standard BRST Quantization

The theory explored in this paper is an SU(3) Yang-Mills gauge field theory, quantized using BRST canonical operator quantization. For convenience, space is assumed to be a three-dimensional grid with $N$ discreet points in a large volume $V$, so that integrals over all of space are replaced by sums over all points, etc. The transition from discrete space back to continuous space will not be addressed in this paper.

The following definitions are used:

- Canonical fields, their momenta: $A_i^a, \Pi_i^a$
- Ghost fields, their momenta: $\eta_a, \mathcal{P}_a$
- Nonminimal fields, their momenta: $A_0^a, \Pi_0^a$
- Anti-ghost fields, their momenta: $\overline{\eta}_a, \overline{\mathcal{P}}_a$
- Gauge constraint: $G_a = -\left(\partial_i \Pi_i^a - g f^{abc} \Pi_i^b A_i^c\right)$
- BRST charge: $\Omega = \sum_x \left[\eta_a G_a + \frac{1}{2} i g f^{abc} \eta_a \eta_b \mathcal{P}_c + \overline{\mathcal{P}}_a \Pi_0^a\right]$
- Ghost number operator: $\mathcal{G} = \frac{1}{2} \sum_x \left( \left[\eta_a, \mathcal{P}_a\right] - \left[\overline{\eta}_a, \overline{\mathcal{P}}_a\right] \right)$, (1)

All of the operators defined above are Hermitian except for the ghost number operator, which is anti-Hermitian. $f^{abc}$ is the SU(3) structure constant, the spatial derivatives $\partial_i$ in the gauge constraints are defined as differences in the usual way for discrete-space analyses.

Canonical quantization is achieved through the following commutators and anticommutators:

$$\begin{aligned}
&\left[\Pi^a_\mu(\vec{x},t), A^b_\nu(\vec{y},t)\right] = -i\delta_{\mu\nu}\delta^{ab}\delta_{xy} \\
&\left\{\eta_a(\vec{x},t), \mathcal{P}_b(\vec{y},t)\right\} = \delta_{ab}\delta_{xy} \\
&\left\{\overline{\eta}_a(\vec{x},t), \overline{\mathcal{P}}_b(\vec{y},t)\right\} = \delta_{ab}\delta_{xy}
\end{aligned} \tag{2}$$

Since calculations are being made in discrete space rather than continuous space, Kroenecker delta functions are used in the above commutators rather than Dirac delta functions. Using these commutation relations, it is straightforward to see that the gauge constraints close in an SU(3) group under commutation and the BRST charge is nilpotent, $\Omega^2 = 0$.

In BRST operator quantization, one requires that any physical states of the theory must be annihilated by the BRST charge. For non-Abelian theories, this is often done in the literature by imposing one of the following two conditions on physical states (see for example [2]):

$$G_a|\phi_\alpha\rangle = A^a_0|\phi_\alpha\rangle = \mathcal{P}_a|\phi_\alpha\rangle = \overline{\mathcal{P}}_a|\phi_\alpha\rangle = 0 \qquad \text{or} \tag{3a}$$

$$\Pi^a_0|\phi'_\alpha\rangle = \eta_a|\phi'_\alpha\rangle = \overline{\eta}_a|\phi'_\alpha\rangle = 0\ . \tag{3b}$$

where the index "$\alpha$" specifies a particular state which meets one of the above two conditions. From (1), it can be seen that both conditions result in states that are BRST-invariant (annihilated by the BRST charge).

A problem with the states in (3) is that they do not have well-defined norms. For example, after expanding the states in (3a) in the Schroedinger representation (and momentum representation in the non-minimal sector), one finds:

$$\langle\phi_\alpha\|\phi_\alpha\rangle = \int\prod_x\left[dA^a_i\, d\Pi^a_0\, d\eta_a\, d\overline{\eta}_a \left|\psi_\alpha\left(A^a_i(\vec{x})\right)\right|^2\right] \tag{4}$$

where $\psi_\alpha\left(A^a_i(\vec{x})\right) = \langle A^a_i(\vec{x})\|\phi_\alpha\rangle$ is the wave function of the physical state with index $\alpha$. Since $G_a|\phi_\alpha\rangle = 0$, the wave functions are independent of the gauge degrees of freedom, so the $dA^a_i(\vec{x})$ produce factors of infinity upon integration over all degrees of freedom (including the gauge degrees of freedom). At the same time, the Berezin integrals over the ghosts and anti-ghosts produce factors of 0. Thus the states defined by the condition (3a) have ill-defined norms. Similar arguments apply to the states defined by (3b).

In [3-7], it was pointed out that BRST-invariant states with finite norms can be constructed if one multiplies certain states by BRST-exact exponentials and quantizes the non-minimal sector with an indefinite metric. Consider the states:

$$\left|\tilde{\phi}_\alpha\right\rangle = \exp\left(\tfrac{1}{2}\left\{\Omega\ ,\sum_x \bar{\eta}_a \chi_a\right\}\right)\left|\phi_\alpha\right\rangle = \exp\left(\tfrac{1}{2}\sum_x \left(\eta_a [G_a, \chi_b]\bar{\eta}_b + \Pi_0^a \chi_a\right)\right)\left|\phi_\alpha\right\rangle \ ,(5)$$

where $\chi_a$ are Hermitian functions of the canonical variables that obey

$$\det[G_a, \chi_b] \neq 0 \ . \tag{6}$$

Since the exponent of (5) is a BRST-exact function, these states are still BRST-invariant. Furthermore, they have well-defined norms given by:

$$\left\langle \tilde{\phi}_\alpha \right\|\left. \tilde{\phi}_\alpha \right\rangle = \int \prod_{x,a} \left[ dA_i^a \det[G_a, \chi_b] \delta(\chi_a) \left|\psi_\alpha\left(A_i^a(\vec{x})\right)\right|^2 \right] = 1 \ . \tag{7}$$

In the above expression, the determinant comes from the first term in the exponent of (5) combined with the Berezin integrals over the ghosts and anti-ghosts. The delta function in (7) comes from the second term in the exponent of (5) and the integrations over $\Pi_0^a$. A subtle requirement in producing this delta function is that the non-minimal momentum states $\left|\Pi_0^a\right\rangle$ must be quantized with indefinite metric. In the context of this quantization, the Hermitian operators $\Pi_0^a$ have imaginary eigenvalues [2,8], leading to the delta function upon integration. The delta function removes the infinite integrals over gauge degrees of freedom, so the states of (5) have finite norms. A different BRST-exact exponential factor in front of the states in eq. (3b) can similarly be used to make those states have finite norms.

It should be noted that both of the classes of states in eq. (3) (as well as their finite-norm counterparts) have zero ghost number. Normally, in addition to imposing the condition that physical states must be BRST-invariant, one also imposes the condition that they have zero ghost number. One reason for this second condition is that since $[\mathcal{G}, \Omega] = \Omega$, any BRST-invariant state can be expressed in terms of a basis of states with definite ghost number. In the context of this basis, it can be shown that the scalar product between two states with definite ghost number vanishes unless the states have opposite ghost number. Therefore, if one wants physical states to have well-defined norms, and one chooses to expand states in a basis with definite ghost number, then only states with ghost number zero can be physical.

## New classes of BRST-invariant states for SU(3)

In the present approach, a different basis of states will be used. In this basis, some of the states do not have well-defined ghost number, but instead are linear combinations of states with different ghost numbers. Construction of the new basis begins by rewriting the BRST charge in terms of "creation" and "destruction" linear combinations of the "off-diagonal" ghosts and ghost momenta:

$$
\begin{aligned}
a_1 &= \tfrac{1}{2}(\eta_1 + i\eta_2 + \mathcal{P}_1 + i\mathcal{P}_2) & b_1 &= \tfrac{1}{2}(\eta_1 + i\eta_2 - \mathcal{P}_1 - i\mathcal{P}_2) \\
a_2 &= \tfrac{1}{2}(\eta_4 - i\eta_5 + \mathcal{P}_4 - i\mathcal{P}_5) & b_2 &= \tfrac{1}{2}(\eta_4 - i\eta_5 - \mathcal{P}_4 + i\mathcal{P}_5) \\
a_3 &= \tfrac{1}{2}(\eta_6 + i\eta_7 + \mathcal{P}_6 + i\mathcal{P}_7) & b_3 &= \tfrac{1}{2}(\eta_6 + i\eta_7 - \mathcal{P}_6 - i\mathcal{P}_7)
\end{aligned} \qquad (8)
$$

The Hermitian conjugates of the above operators will be denoted by $a_i^+$ and $b_i^+$. The indices on the ghost fields on the right-hand sides of the above equations assume a basis for SU(3) defined by the standard 8 Gell-Mann 3x3 matrices. Similar creation and destruction operator definitions are obtained for the anti-ghost fields by putting bars over each of the operators in (8). With these definitions, the commutation relations between these operators and their complex conjugates are:

$$
\begin{aligned}
\{a_i(\vec{x},t), a_j^+(\vec{y},t)\} &= \delta_{ij}\delta_{xy} & \{b_i(\vec{x},t), b_j^+(\vec{y},t)\} &= -\delta_{ij}\delta_{xy} \\
\{\bar{a}_i(\vec{x},t), \bar{a}_j^+(\vec{y},t)\} &= \delta_{ij}\delta_{xy} & \{\bar{b}_i(\vec{x},t), \bar{b}_j^+(\vec{y},t)\} &= -\delta_{ij}\delta_{xy}\ ,
\end{aligned} \qquad (9)
$$

and all other anti-commutators vanish. The $a_i^+$ operators and their complex conjugates behave like standard creation and destruction operators, while the $b_i^+$ operators have the "wrong" sign.

Similar linear combinations are defined for the following "off-diagonal" operators:

$$
\begin{aligned}
G_1^\pm &= \tfrac{1}{2}(G_1 \mp iG_2) & \Pi_1^\pm &= \tfrac{1}{2}(\Pi_0^1 \mp i\Pi_0^2) & A_1^\pm &= (A_0^1 \mp iA_0^2) \\
G_2^\pm &= \tfrac{1}{2}(G_4 \pm iG_5) & \Pi_2^\pm &= \tfrac{1}{2}(\Pi_0^4 \pm i\Pi_0^5) & A_2^\pm &= (A_0^4 \pm iA_0^5) \\
G_3^\pm &= \tfrac{1}{2}(G_6 \mp iG_7) & \Pi_3^\pm &= \tfrac{1}{2}(\Pi_0^6 \mp i\Pi_0^7). & A_3^\pm &= (A_0^6 \mp iA_0^7)
\end{aligned} \qquad (10)
$$

It should be noted that the third "off-diagonal" constraints like $G_3^\pm$ are distinct from the constraint $G_3$ that is in the diagonal "3" gauge direction.

With these definitions, the BRST charge can be rewritten:

$$\Omega = \sum_x [\Omega_1(\vec{x}) + \Omega_2(\vec{x}) + \Omega_3(\vec{x})]$$

$$\Omega_1 = (a_i + b_i)G_i^+ + (a_i^+ + b_i^+)G_i^- + \eta_3 G_3 + \eta_8 G_8$$

$$\Omega_2 = -\tfrac{1}{8} g\varepsilon^{ijk} \left[ \left(a_i^+ + b_i^+\right)\left(a_j^+ a_k^+ - b_j^+ b_k^+\right) + \left(a_k a_j - b_k b_j\right)\left(a_i + b_i\right)\right]$$

$$+\tfrac{1}{4} g \left[ 2\eta_3\left(a_1^+ a_1 - b_1^+ b_1\right) - \left(\eta_3 + \sqrt{3}\eta_8\right)\left(a_2^+ a_2 - b_2^+ b_2\right) - \left(\eta_3 - \sqrt{3}\eta_8\right)\left(a_3^+ a_3 - b_3^+ b_3\right)\right]$$

$$+\tfrac{1}{4} g \left[ 2\mathcal{P}_3\left(a_1^+ + b_1^+\right)\left(a_1 + b_1\right) - \left(\mathcal{P}_3 + \sqrt{3}\mathcal{P}_8\right)\left(a_2^+ + b_2^+\right)\left(a_2 + b_2\right) - \left(\mathcal{P}_3 - \sqrt{3}\mathcal{P}_8\right)\left(a_3^+ + b_3^+\right)\left(a_3 + b_3\right)\right]$$

$$\Omega_3 = \left(\overline{a}_i - \overline{b}_i\right)\Pi_i^+ + \left(\overline{a}_i^+ - \overline{b}_i^+\right)\Pi_i^- + \overline{\mathcal{P}}_3\Pi_0^3 + \overline{\mathcal{P}}_8\Pi_0^8 \ . \quad (11)$$

This notation highlights the threefold symmetry of the "off-diagonal" generators of SU(3), especially in $\Omega_2$ , where the first term has a factor of $\varepsilon^{ijk}$ . In this notation, the ghost number operator becomes:

$$\mathcal{G} = \tfrac{1}{2}\sum_x \left(b_i^+ a_i - a_i^+ b_i - \overline{b}_i^+ \overline{a}_i + \overline{a}_i^+ \overline{b}_i + \left[\eta_3, \mathcal{P}_3\right] - \left[\overline{\eta}_3, \overline{\mathcal{P}}_3\right] + \left[\eta_8, \mathcal{P}_8\right] - \left[\overline{\eta}_8, \overline{\mathcal{P}}_8\right]\right). \quad (12)$$

The BRST-invariant states that will be constructed here feature complete decoupling between the ghost, anti-ghost, non-minimal, and minimal sectors. The ghost sector will be constructed first, and to do that, it is necessary to define a ghost vacuum $\left|0_{gh}\right\rangle$ that satisfies:

$$a_i\left|0_{gh}\right\rangle = b_i\left|0_{gh}\right\rangle = \overline{a}_i\left|0_{gh}\right\rangle = \overline{b}_i\left|0_{gh}\right\rangle = \mathcal{P}_3\left|0_{gh}\right\rangle = \mathcal{P}_8\left|0_{gh}\right\rangle = \overline{\mathcal{P}}_3\left|0_{gh}\right\rangle = \overline{\mathcal{P}}_8\left|0_{gh}\right\rangle = 0 . \quad (13)$$

From (12), it can be seen that this ghost vacuum has zero ghost number. Nontrivial BRST-invariant states can be created by acting on this vacuum with off-diagonal creation operators and diagonal ghosts.

In particular, consider the following operators:

$$\xi_1^+ = a_1^+ b_1^+ \eta_3 \eta_8 \overline{\eta}_3 \overline{\eta}_8$$

$$\xi_2^+ = a_2^+ b_2^+ a_3^+ b_3^+ \eta_3 \eta_8 \overline{\eta}_3 \overline{\eta}_8$$

$$\xi_3^+ = a_2^+ b_2^+ \eta_3 \eta_8 \overline{\eta}_3 \overline{\eta}_8$$

$$\xi_4^+ = a_3^+ b_3^+ a_1^+ b_1^+ \eta_3 \eta_8 \overline{\eta}_3 \overline{\eta}_8$$

$$\xi_5^+ = a_3^+ b_3^+ \eta_3 \eta_8 \overline{\eta}_3 \overline{\eta}_8$$

$$\xi_6^+ = a_1^+ b_1^+ a_2^+ b_2^+ \eta_3 \eta_8 \overline{\eta}_3 \overline{\eta}_8$$

$$\xi_7^+ = \frac{1}{\sqrt{2}}\left(b_1^+ a_2^+ a_3^+ - b_2^+ a_3^+ a_1^+\right)$$

$$\xi_8^+ = \frac{1}{\sqrt{2}}\left(a_1^+ b_2^+ b_3^+ - a_2^+ b_3^+ b_1^+\right)$$

$$\xi_9^+ = \frac{1}{\sqrt{6}}\left(b_1^+ a_2^+ a_3^+ + b_2^+ a_3^+ a_1^+ - 2b_3^+ a_1^+ a_2^+\right)$$

$$\xi_{10}^+ = \frac{1}{\sqrt{6}}\left(a_1^+ b_2^+ b_3^+ + a_2^+ b_3^+ b_1^+ - 2a_3^+ b_1^+ b_2^+\right)$$

$$\xi_{11}^+ = \frac{1}{2}\left(b_1^+ b_2^+ b_3^+ + b_1^+ a_2^+ a_3^+ + b_2^+ a_3^+ a_1^+ + b_3^+ a_1^+ a_2^+\right)$$

$$\xi_{12}^+ = \frac{1}{2}\left(a_1^+ a_2^+ a_3^+ + a_1^+ b_2^+ b_3^+ + a_2^+ b_3^+ b_1^+ + a_3^+ b_1^+ b_2^+\right) \tag{14}$$

The Hermitian conjugates of these operators will be denoted by $\xi_n^-$. Acting on the vacuum, these operators form states that are annihilated by the non-Abelian ghost part of the BRST charge:

$$\Omega_2(\vec{x})\xi_n^+(\vec{x})\left|0_{gh}\right\rangle = 0\,. \tag{15}$$

The states are normalized to either $\pm 1$ in the following sense:

$$\left\langle 0_{gh}\right|\xi_n^-(\vec{x})\left(\overline{P}_8\overline{P}_3 P_8 P_3 + \overline{\eta}_8\overline{\eta}_3\eta_8\eta_3\right)\xi_m^+(\vec{y})\left|0_{gh}\right\rangle = (-1)^n\,\delta_{nm}\delta_{xy} \tag{16}$$

The first six operators which are grouped into doublet pairs that have zero ghost number, while the second six operators are grouped into triplets that do not have well-defined ghost numbers.

Among the SU(N) groups, only SU(3) has ghost operators like those in (14) that are annihilated by the non-Abelian ghost part of the BRST charge. The trick is to find states that are annihilated both by the terms in $\Omega_2$ involving 3 creation operators and by those involving 3 destruction operators. In SU(3), this is possible due to the fact that there are exactly 3 sets of off-diagonal creation and destruction operators. As a result, when the terms in $\Omega_2$ involving 3 creation operators act on the triplet states, they form terms with "full" states (involving every creation operator) that have opposite signs, so they cancel. Conversely, when the terms in $\Omega_2$ involving 3 destruction operators act on the triplets, they form "empty" states that cancel. In any other SU(N) group in which off-diagonal elements have been turned into creation and destruction operators, it would not be possible to form both "full" and "empty" states by the action of $\Omega_2$, so other SU(N) groups do not have analogous solutions to (15). The doublet states are similarly unique to SU(3).

Once (15) has been satisfied by the use of one of the operators in (14), it is straightforward to form states that are annihilated by the remaining parts of the BRST charge. To do this, one may define "matter" states $\left|M_n^\alpha\right\rangle$ made from the minimal canonical variables. The index "*n*" on each of these states corresponds to one of the ghost operators in (14), and for each *n*, the index $\alpha$ identifies distinct states that satisfy the conditions:

$$G_1^+\left|M_1^\alpha\right\rangle = G_2^-\left|M_1^\alpha\right\rangle = G_3^-\left|M_1^\alpha\right\rangle = 0$$

$$G_1^-\left|M_2^\alpha\right\rangle = G_2^+\left|M_2^\alpha\right\rangle = G_3^+\left|M_2^\alpha\right\rangle = 0$$

$$G_1^-\left|M_3^\alpha\right\rangle = G_2^+\left|M_3^\alpha\right\rangle = G_3^-\left|M_3^\alpha\right\rangle = 0$$

$$G_1^+\left|M_4^\alpha\right\rangle = G_2^-\left|M_4^\alpha\right\rangle = G_3^+\left|M_4^\alpha\right\rangle = 0$$

$$G_1^-\left|M_5^\alpha\right\rangle = G_2^-\left|M_5^\alpha\right\rangle = G_3^+\left|M_5^\alpha\right\rangle = 0$$

$$G_1^+\left|M_6^\alpha\right\rangle = G_2^+\left|M_6^\alpha\right\rangle = G_3^-\left|M_6^\alpha\right\rangle = 0$$

$$G_a\left|M_n^\alpha\right\rangle = 0 \quad \text{for} \quad n > 6 \tag{17}$$

Because of the spatial derivatives in the constraints, the states obeying (17) mix different points of space, so one cannot define the matter states independently at each point of space. Nonetheless, over some region of space, one may build states like $\prod_x \xi_n^+(\vec{x})\left|0_{gh}\right\rangle\left|M_n^\alpha\right\rangle$ that are annihilated by $\sum_x(\Omega_1 + \Omega_2)$.

To complete the definition of BRST-invariant states, one must define states $\left|\Lambda_{n,p}\right\rangle$ in the non-minimal sector which, together with the anti-ghosts, are annihilated by $\Omega_3$. For each ghost operator in (14), the diagonal anti-ghost structure is also specified. This is done to ensure that states with finite norms can be constructed (see below). Since the first six operators in (14) all have a factor of $\bar{\eta}_3\bar{\eta}_8$ in them, one must impose the following restrictions on the non-minimal states in order to ensure annihilation by $\Omega_3$:

$$\Pi_0^3\left|\Lambda_{n,p}\right\rangle = \Pi_0^8\left|\Lambda_{n,p}\right\rangle = 0 \quad \text{for} \quad n \le 6 \tag{18}$$

There is much more freedom in the off-diagonal anti-ghosts, but whatever anti-ghost structure is chosen, the following corresponding restrictions must be made on the non-minimal states:

$$\Omega_3(\vec{x})\xi_n^+(\vec{x})\bar{\chi}_p^+(\vec{x})\left|0_{gh}\right\rangle\left|\Lambda_{n,p}(\vec{x})\right\rangle = 0, \tag{19}$$

where the operators $\bar{\chi}_p^+$ are formed from combinations of off-diagonal anti-ghost creation operators. In the above equation, explicit $\vec{x}$ dependence has been shown for the non-minimal states. This highlights the fact that since $\Omega_3$ does not have any spatial derivatives in it, independent state bases satisfying (19) can be constructed at each point of space.

Some specific examples of off-diagonal ghost operators and non-minimal state restrictions that obey (19) are:

$$\bar{\chi}_0^+ = 1 \qquad \Pi_i^- \left|\Lambda_{n,0}\right\rangle = 0$$

$$\bar{\chi}_1^+ = \bar{a}_1^+ \bar{b}_1^+ \qquad \Pi_1^+ \left|\Lambda_{n,1}\right\rangle = \Pi_2^- \left|\Lambda_{n,1}\right\rangle = \Pi_3^- \left|\Lambda_{n,1}\right\rangle = 0$$

$$\bar{\chi}_2^+ = \bar{a}_2^+ \bar{b}_2^+ \bar{a}_3^+ \bar{b}_3^+ \qquad \Pi_1^- \left|\Lambda_{n,2}\right\rangle = \Pi_2^+ \left|\Lambda_{n,2}\right\rangle = \Pi_3^+ \left|\Lambda_{n,2}\right\rangle = 0$$

$$\bar{\chi}_3^+ = \bar{a}_2^+ \bar{b}_2^+ \qquad \Pi_1^- \left|\Lambda_{n,3}\right\rangle = \Pi_2^+ \left|\Lambda_{n,3}\right\rangle = \Pi_3^- \left|\Lambda_{n,3}\right\rangle = 0$$

$$\bar{\chi}_4^+ = \bar{a}_3^+ \bar{b}_3^+ \bar{a}_1^+ \bar{b}_1^+ \qquad \Pi_1^+ \left|\Lambda_{n,4}\right\rangle = \Pi_2^- \left|\Lambda_{n,4}\right\rangle = \Pi_3^+ \left|\Lambda_{n,4}\right\rangle = 0$$

$$\bar{\chi}_5^+ = \bar{a}_3^+ \bar{b}_3^+ \qquad \Pi_1^- \left|\Lambda_{n,5}\right\rangle = \Pi_2^- \left|\Lambda_{n,5}\right\rangle = \Pi_3^+ \left|\Lambda_{n,5}\right\rangle = 0$$

$$\bar{\chi}_6^+ = \bar{a}_1^+ \bar{b}_1^+ \bar{a}_2^+ \bar{b}_2^+ \qquad \Pi_1^+ \left|\Lambda_{n,6}\right\rangle = \Pi_2^+ \left|\Lambda_{n,6}\right\rangle = \Pi_3^- \left|\Lambda_{n,6}\right\rangle = 0$$

$$\bar{\chi}_7^+ = \bar{a}_1^+ \bar{b}_1^+ \bar{a}_2^+ \bar{b}_2^+ \bar{a}_3^+ \bar{b}_3^+ \qquad \Pi_i^+ \left|\Lambda_{n,7}\right\rangle = 0 \tag{20}$$

The above examples are useful in the discussions below about creating states with well-defined norms.

Putting everything together, one can construct the following BRST-invariant states:

$$\left|\varphi_{N,P}^{\alpha}\right\rangle = \prod_x \left[ \exp\left(\tfrac{1}{2}\left\{\Omega \ , K_{n(x)}(\vec{x})\right\}\right) \xi_{n(x)}^+(\vec{x}) \bar{\chi}_{p(x)}^+(\vec{x}) \left|\Lambda_{n(x),p(x)}(\vec{x})\right\rangle \left|M_{n(x)}^{\alpha}\right\rangle \ \right]_x \left|0_{gh}\right\rangle . \tag{21}$$

In the above states, the ghost ($n$) and off-diagonal-anti-ghost ($p$) structure is allowed to be different at different points of space, and the indices $N$ and $P$ are used to denote some unique combination of $n$'s and $p$'s at different points. As mentioned above, due to the spatial derivatives in the gauge constraints and the different constraint conditions (17) for different ghost states, the ghost structure ($n$) may have to remain constant over finite regions of space.

Just as in (5), a BRST-exact exponential factor has been introduced in (21) in order to create states with finite norms. The gauge-fixing fermion $K_n$ can be given by:

$$\text{for } n \le 6, \qquad K_n \equiv P_3 A_0^3 + P_8 A_0^8 + \bar{\eta}_{\bar{a}} \chi_{\bar{a}} \qquad \bar{a} \in (1,2,4,5,6,7)$$

$$\text{for } n > 6, \qquad K_n \equiv \bar{\eta}_a \chi_a \ , \tag{22}$$

where $\chi_a$ are functions of the canonical variables that obey $\det[G_a , \chi_b] \ \neq 0$ . To understand the norms of these states, it is helpful to expand the BRST-exact exponentials. For example, for $n \le 6$, one finds:

$$\left\{\Omega \ , K_n\right\} = -i\bar{P}_3 P_3 - i\bar{P}_8 P_8 + A_0^3\left(G_3 + igf^{3bc}\eta_b P_c\right) + A_0^8\left(G_8 + igf^{8bc}\eta_b P_c\right) + \chi_{\bar{a}} \Pi_0^{\bar{a}}$$
$$+ \eta_a \left[G_a , \chi_{\bar{b}}\right] \ \bar{\eta}_{\bar{b}} \tag{23}$$

The first two terms above provide the ghost factor needed for normalization as in (16), while the other terms on the first line produce delta functions that remove the infinite integrals over gauge degrees of freedom.

$$\left\langle \varphi^{\alpha}_{n,p} \middle\| \varphi^{\alpha}_{n,p} \right\rangle = \int \prod_x \left[ dA_i^a \delta(\chi_{\bar{a}}) \delta(G_3 - c_{3n}) \delta(G_8 - c_{8n}) \left| \psi_n^{\alpha}\left(A_i^a(\vec{x})\right)\right|^2 \right] = \pm 1 \quad (24)$$

The constant $c_{3n}$ comes from the action of the operator $igf^{3bc}\eta_b P_c$ on the ghost parts of the physical states. The physical states are eigenstates of this operator (and its "8" counterpart), returning constants. For example, $c_{31} = g$.

In obtaining (24), the off-diagonal anti-ghost structure of the states has implicitly been chosen so that the term in the second line of (23) does not contribute. Any of the examples in (20) will suffice for this purpose, since all of them satisfy:

$$\left\langle 0_{gh} \left| \overline{\chi}_q^-(\vec{x}) \overline{\eta}_{\bar{a}} \overline{\chi}_p^+(\vec{y}) \right| 0_{gh} \right\rangle = 0 . \quad (25)$$

The vanishing of the contribution from the second term in (23) should be compared to (7) in which a similar term produces a Fadeev-Popov determinant.

An obvious problem with (24) is the fact that some of the states have negative norms. In order to preserve unitarity, these states must be removed. To do this, one can introduce the following negative-norm counting operators:

$$\begin{aligned}
N_1^- &= -\exp\left(\tfrac{1}{2}\{\Omega\ , K_n\}\right) a_1^+ b_1^+ b_1 a_1 \exp\left(-\tfrac{1}{2}\{\Omega\ , K_n\}\right) \\
N_2^- &= -\exp\left(\tfrac{1}{2}\{\Omega\ , K_n\}\right) a_2^+ b_2^+ b_2 a_2 \exp\left(-\tfrac{1}{2}\{\Omega\ , K_n\}\right) \\
N_3^- &= -\exp\left(\tfrac{1}{2}\{\Omega\ , K_n\}\right) a_3^+ b_3^+ b_3 a_3 \exp\left(-\tfrac{1}{2}\{\Omega\ , K_n\}\right) \\
N_T^- &= -\exp\left(\tfrac{1}{2}\{\Omega\ , K_n\}\right) \left[ b_1^+ b_2^+ b_3^+ b_3 b_2 b_1 + b_1^+ a_2^+ a_3^+ a_3 a_2 b_1 \right. \\
&\qquad \left. + b_2^+ a_3^+ a_1^+ a_1 a_3 b_2 + b_3^+ a_1^+ a_2^+ a_2 a_1 b_3 \right] \exp\left(-\tfrac{1}{2}\{\Omega\ , K_n\}\right) \\
\overline{N}_i^- &= -\exp\left(\tfrac{1}{2}\{\Omega\ , K_n\}\right) \overline{b}_i^+ \overline{b}_i \exp\left(-\tfrac{1}{2}\{\Omega\ , K_n\}\right).
\end{aligned} \quad (26)$$

Each local ghost and anti-ghost state constructed above is an eigenstate of all of the above operators. As a result, one can define a "physical" subset of the states in (21) that satisfy the condition:

$$\sum_x \left(N_i^- + N_T^- + \overline{N}_i^-\right) \left| \tilde{\varphi}^{\alpha}_{n,p} \right\rangle = 2m \left| \tilde{\varphi}^{\alpha}_{n,p} \right\rangle, \text{ where } m \text{ is an integer.} \quad (27)$$

States defined in this way are orthonormal and preserve unitarity:

$$\left\langle \tilde{\varphi}^{\alpha}_{n,p} \middle\| \tilde{\varphi}^{\alpha}_{n,p} \right\rangle = 1 \quad (28)$$

As an aside, it is interesting to point out two alternative ways to restrict the space of physical states to preserve unitarity. The first is to demand that every state must be an eigenstate of a simpler negative counting operator

$$N^{-} = -\exp\left(\tfrac{1}{2}\{\Omega\ , K_n\}\right)\left(b_i^{+} b_i + \bar{b}_i^{+} \bar{b}_i\right)\exp\left(-\tfrac{1}{2}\{\Omega\ , K_n\}\right),$$

and that physical states must have even eigenvalues. This choice is more restrictive than the approach presented above because it eliminates the ghost operators $\xi_{11}^{+}$ and $\xi_{12}^{+}$ which do not create eigenstates of $N^{-}$. These two ghost operators are interesting to keep in the space of states, since they are the only ghost operators that preserve the symmetry of the off-diagonal elements of SU(3). A second alternative to restrict the space of states is to demand that physical states must have ghost number zero and also be anti-BRST invariant [9]. These restrictions cause the anti-ghost structure of the states to be the same as the ghost structure, leading to states of the form:

$$\prod_x \left[ \exp\left(\tfrac{1}{2}\{\Omega\ , K_n(\vec{x})\}\right)\xi_n^{+}(\vec{x})\chi_n^{+}(\vec{x})\left|\Lambda_{n,n}(\vec{x})\right\rangle \right]_x \left|0_{gh}\right\rangle\left|M_n^{\alpha}\right\rangle\ ,\ n \leq 6$$

This prescription is closer to standard approaches, since it requires zero ghost number, but it removes all of the ghost triplets from physical states.

## Summary

New classes of BRST-invariant states have been explicitly constructed for SU(3) gauge field theory. It has been argued that these classes of states only appear in SU(3) and not in other SU(N) field theories. Since the states have well-defined, positive norms, there seems to be nothing stopping one from interpreting all of these states as “physical”. This leads one to wonder whether SU(3) gauge theories might feature some kind of ghost excitations, perhaps over finite regions of space.

Certainly, if these states are physical, they imply the possibility of a much more complicated ghost structure for QCD than previously considered. Is it possible that ghost excitations could shed some light on the confinement problem? It is certainly intriguing that the ghost states presented here only appear in triplets or in positive-norm-negative-norm pairs, while in QCD (at least at low temperatures), quarks only appear in triplets or quark-anti-quark pairs. It would be interesting to see how close of an analogy could be drawn between ghosts and quarks.